\documentclass[preprint,3p,number,sort&compress]{elsarticle}

\usepackage{amssymb}
\usepackage{amsthm}
\usepackage{lineno}
\usepackage{amsmath}
\usepackage{microtype}
\usepackage[american]{babel}
\usepackage{booktabs}
\usepackage{graphicx}
\usepackage{epstopdf}
\usepackage{caption}
\usepackage{float}
\usepackage{subfigure}
\usepackage{algorithm}
\usepackage{algorithmic}
\usepackage{epstopdf}
\usepackage{url}
\usepackage{xcolor} 
\usepackage{lmodern}
\usepackage{microtype}
\usepackage{booktabs}    
\usepackage{multirow}    
\usepackage{multicol}

\DeclareCaptionLabelFormat{figformat}{Fig.\thefigure} 
\begin{document}
\begin{frontmatter}

\title{\bfseries Measuring node similarity using minimum cycles in networks
\tnoteref{t1}}
\tnotetext[t1]{This work was supported by Yunnan Fundamental Research Projects (grant NO. 202401AT070359)}

\author[label1,label2,label3]{Bo Yang\corref{cor1}}
 \ead{yangbo@kust.edu.cn}
 \cortext[cor1]{Corresponding author}
\affiliation[label1]{organization={Data Science Research Center},
             addressline={Kunming University of Science and Technology},
             city={Kunming},
             postcode={650500},
             state={Yunnan},
             country={China}}

\affiliation[label2]{organization={Faculty of Science},
             addressline={Kunming University of Science and Technology},
             city={Kunming},
             postcode={650500},
             state={Yunnan},
             country={China}}

\affiliation[label3]{organization={Yunnan Key Laboratory of Complex Systems and Brain-Inspired Intelligence},
             addressline={Kunming University of Science and Technology},
             city={Kunming},
             postcode={650500},
             state={Yunnan},
             country={China}}

\begin{abstract}
Cycles are ubiquitous in various networks such as social, biological, and technological systems, where they play a significant functional and dynamical role. This paper proposes a node similarity measure based on minimal simple cycles, referred to as cycle similarity. Specifically, the metric quantifies the similarity between two nodes by considering the minimal cycles that connect them through their neighboring nodes, with an upper bound imposed on the cycle size to ensure computational feasibility. We then systematically examine the effectiveness and applicability of this similarity measure through two fundamental tasks: link prediction and community detection. To address the scarcity of cycles in link prediction, an edge-addition correction strategy is introduced, whereby the existence of a candidate edge is hypothetically assumed before computing node similarity. Experimental results demonstrate that this correction leads to improved performance on datasets including karate, INT, PPI, and Grid. In hierarchical community detection using cycle similarity, we find that the significance of cyclic structures (reflected by Z-scores), the presence of pendant nodes with degree one, and the existence of cut vertices are the primary factors influencing the algorithm's performance.
\end{abstract}

\end{frontmatter}

\section{Introduction}\label{Indro}
Node similarity is a fundamental concept in network science, serving as the cornerstone for numerous analytical tasks including link prediction\cite{PA2011,EPJB2009,ACM2016}, community detection\cite{PR2010,PR2016,PA2022,PNAS2002,NS2022}, and recommendation systems\cite{PR2012}. It quantifies the degree of relatedness or affinity between two nodes based on their structural positions within the network. Traditional approaches to measuring node similarity often rely on local neighborhood information, such as common neighbors, Jaccard coefficient, Adamic-Adar index, or global structural properties like shortest path length and PageRank\cite{PA2011,EPJB2009}. These methods effectively capture various aspects of node proximity but primarily focus on direct connections or linear paths between nodes, thereby providing a partial view of the complex interdependencies that exist in real-world networks.

Cycles, or closed walks in a network where the starting and ending nodes are identical, are ubiquitous structural motifs across diverse types of networks, including social networks, biological networks, technological networks, and information networks\cite{Entropy2025,CP2021,CAAI2025}. Their widespread presence is not merely coincidental but reflects inherent organizational principles and functional constraints of these systems. Cycles play a pivotal role in network dynamics and functionality: in social networks, they represent closed communication loops or collaborative groups; in metabolic networks, they correspond to cyclic biochemical reactions; and in transportation networks, they signify round trips or feedback mechanisms. Despite their prevalence and importance, the explicit incorporation of cycle structures into node similarity measures has received relatively limited attention compared to other structural features. Specifically, the application of cycles to enhance node similarity assessment remains an underexplored area. Given the rich structural information encoded in cycles, introducing minimal cycles as a descriptor for node similarity presents a valuable research direction that could deepen our understanding of node relationships and improve the performance of network analysis tasks.

Based on this insight, this paper proposes a novel node similarity metric based on minimal cycles, termed cycle similarity. The metric quantifies the similarity between two nodes by considering the minimal cycles that connect them through their neighboring nodes, with an upper bound imposed on the cycle size to ensure computational feasibility. By focusing on minimal cycles, we aim to capture the tightest structural relationships between nodes while avoiding the inclusion of spurious long-range dependencies that may dilute the similarity signal.

To validate the effectiveness of the proposed cycle similarity metric, we conduct comprehensive experiments on two critical network analysis tasks: link prediction and community detection. For the link prediction task, we first analyze several real-world networks, evaluating performance using three widely adopted metrics: Area Under the Receiver Operating Characteristic Curve (AUC), Precision, and Ranking Score. These metrics collectively assess both the ability to correctly identify true links and the quality of ranking potential links. Recognizing that many real-world networks contain a significant number of degree-one pendant nodes and tree-like structures, which may pose challenges for cycle-based metrics due to the absence of natural cycles, we further propose an edge-augmented correction for cycle similarity. This correction assumes the existence of a virtual edge between the two nodes whose similarity is being computed before calculating the cycle similarity, thereby enabling meaningful comparisons even in sparse or acyclic substructures. For community detection, we introduce a Cycle-based Hierarchical Clustering (CHC) algorithm that leverages the proposed cycle similarity to guide the agglomerative clustering process. We evaluate the CHC algorithm's performance under different network structures using two standard metrics: Fraction of Vertices Identified Correctly (FVIC), which measures the proportion of nodes assigned to their correct communities, and Normalized Mutual Information (NMI), which quantifies the overlap between detected communities and ground-truth communities. Through these systematic evaluations, we aim to demonstrate that cycle similarity outperforms existing state-of-the-art metrics in capturing nuanced structural relationships within networks.

The remainder of this paper is organized as follows. In Sect.~\ref{S2}, we first define the concept of cycle similarity and introduce commonly used evaluation metrics for link prediction. Subsequently, we detail the hierarchical clustering community detection algorithm based on cycle similarity, along with its associated observation indicators. In Sect.~\ref{S3}, we conduct link prediction and community detection experiments on several real-world networks, followed by an in-depth analysis and discussion of the results. Finally, Sect.~\ref{S4} summarizes the research findings and conclusions, and outlines future research directions. \par

\section{Model}\label{S2}
\subsection{Cycle similarity}
  
In an undirected, unweighted simple graph \( G = (V, E) \), where $V$ is the vertex set and $E$ is the edge set, a simple cycle is defined as a vertex sequence \( v_1 \to v_2 \to \dots \to v_k \to v_1 \) where \( v_1, \dots, v_k \) are distinct (with \( k \geq 3 \)), each edge \( (v_i, v_{i+1}) \) for \( i=1,\dots,k-1 \) and \( (v_k, v_1) \) are distinct edges in \( E \), and no self-loops exist (i.e., \( v_i \neq v_{i+1} \)). The minimum cycle is the simple cycle with the smallest length among all simple cycles, where the length is defined as the number of edges it contains (e.g., a triangular cycle has length 3).

This paper proposes a node similarity metric based on minimal cycles, termed cycle similarity. For any two nodes $x$ and $y$, their similarity 
$s_{xy}$ is defined as:

\begin{eqnarray}
s_{xy}= \sum_{i \in \Gamma(x)} \frac{1}{L^c_{x,i,y}} + \sum_{j \in \Gamma(y)} \frac{1}{L^c_{y,j,x}}, \label{e1}
\end{eqnarray}
\noindent
where \( \Gamma(x) \) denotes the neighbor set of node \( x \), \( L^c_{x,i,y} \) represents the length of the minimal cycle containing nodes \( x \), \( i \), and \( y \) (with \( i \neq y \)), \( L^c_{y,j,x} \) is defined analogously for node \( y \). A minimal cycle is the shortest simple cycle that includes the three specified nodes. For instance, if nodes \( x \), \( i \), and \( y \) form a triangle, then \( L^c_{x,i,y} = 3 \). If they do not form a triangle but lie on a quadrilateral, then \( L^c_{x,i,y} = 4 \), and so forth.

Due to the high computational complexity of enumerating all minimal cycles, we restrict the cycle size to a maximum value \( L_{\max}^c \). Specifically, only cycles with length not exceeding \( L_{\max}^c \) are considered in the similarity calculation. For example, setting \( L_{\max}^c = 5 \) implies that only triangles, quadrilaterals, and pentagons formed by \( x \), \( i \), and \( y \) are taken into account, while larger cycles are disregarded.
 
This restriction significantly reduces the computational overhead while preserving the local structural information essential for similarity assessment. The parameter \( L_{\max}^c \) can be adjusted based on the specific requirements of the application and the computational resources available.

\begin{multicols}{2}
  
  \begin{figure}[H]
    \centering
    \includegraphics[width=0.8\linewidth]{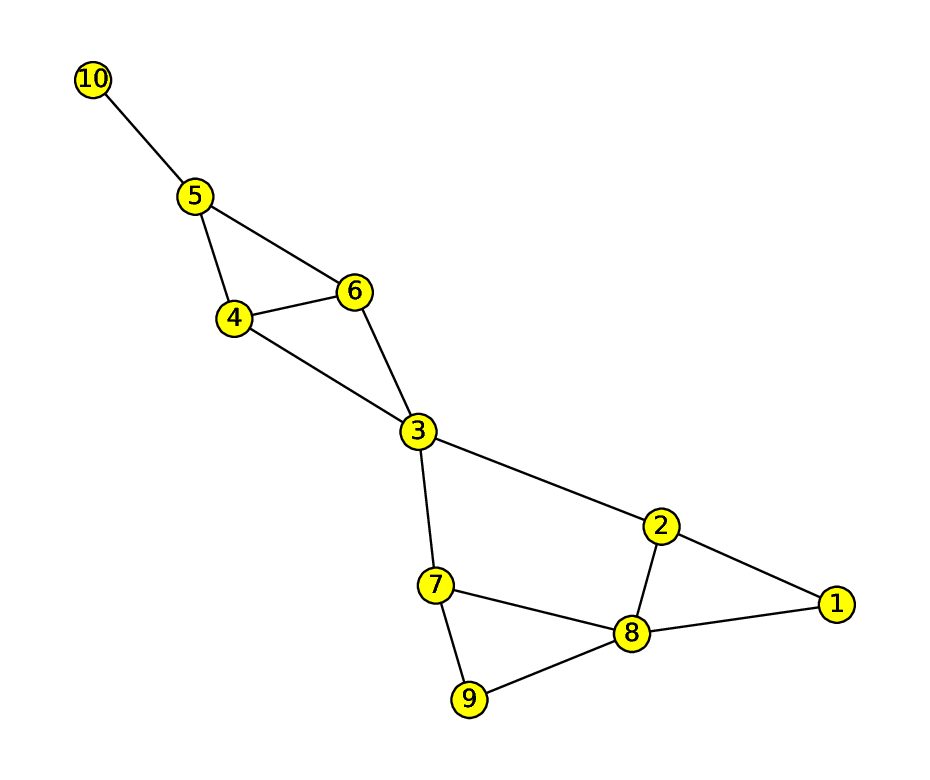}
    \caption{Schematic representation of the example network used to demonstrate the pairwise cycle similarity computation.}
    \label{fig1}
  \end{figure}
  
  \columnbreak  
  
  \begin{table}[H]
    \centering
    \caption{All simple cycles identified in the example network.}
    \label{tab1}
    \begin{tabular}{cl}
      \toprule 
      Cycle Size & Corresponding Cycles \\
      \midrule 
      $L^c = 3$ & [1, 2, 8, 1] \\
                & [3, 4, 6, 3]\\
                & [4, 5, 6, 4] \\
                & [7, 8, 9, 7]\\
      \midrule 
      $L^c = 4$ & [2, 3, 7, 8, 2] \\
                & [3, 4, 5, 6, 3] \\
      \midrule 
      $L^c = 5$ & [1, 2, 3, 7, 8, 1] \\
                & [2, 3, 7, 9, 8, 2] \\
      \midrule 
      $L^c = 6$ & [1, 2, 3, 7, 9, 8, 1] \\
      \bottomrule 
    \end{tabular}
  \end{table}
  
\end{multicols}

To provide an intuitive illustration of the above definition, we employ an exemplary network comprising 10 nodes and 14 edges in Figure \ref{fig1}. Table \ref{tab1} enumerates all simple cycles present in this network. Table \ref{tab2} reports the pairwise cycle similarity values computed for two types of node pairs: those with existing edges (relevant for community detection) and those without edges (relevant for link prediction). Node pairs with zero similarity are omitted from the table.

\begin{table}[ht]
\centering 
\caption{Pairwise cycle similarity scores for node pairs with and without edges in the example network.}
\label{tab2}
\begin{tabular}{cccc|ccc}
\toprule
\multicolumn{4}{c|}{Existing Edges} & \multicolumn{3}{c}{Non-existing Edges} \\
\midrule
\multicolumn{1}{c}{$L^c_{max}=3$} & \multicolumn{1}{c}{$L^c_{max}=4$} & \multicolumn{1}{c}{$L^c_{max}=5$} & \multicolumn{1}{c|}{$L^c_{max}\geq6$} & 
\multicolumn{1}{c}{$L^c_{max}=4$} & \multicolumn{1}{c}{$L^c_{max}=5$} & \multicolumn{1}{c}{$L^c_{max}\geq6$} \\
\midrule 
(4, 6), 1.33 & (4, 6), 1.33 & (2, 8), 1.37 & (2, 8), 1.37 & (2, 7), 1.00 & (2, 7), 1.40 & (2, 7), 1.40 \\
(1, 2), 0.67 & (2, 8), 1.17 & (7, 8), 1.37 & (7, 8), 1.37 & (3, 8), 1.00 & (3, 8), 1.40 & (3, 8), 1.40 \\
(1, 8), 0.67 & (7, 8), 1.17 & (4, 6), 1.33 & (4, 6), 1.33 & (3, 5), 1.00 & (3, 5), 1.00 & (3, 5), 1.00 \\
(2, 8), 0.67 & (3, 4), 0.92 & (3, 4), 0.92 & (1, 8), 1.03 &  & (1, 3), 0.80 & (1, 7), 0.97 \\
(3, 4), 0.67 & (3, 6), 0.92 & (3, 6), 0.92 & (8, 9), 1.03 &  & (1, 7), 0.80 & (2, 9), 0.97 \\
(3, 6), 0.67 & (4, 5), 0.92 & (4, 5), 0.92 & (3, 4), 0.92 &  & (2, 9), 0.80 & (1, 3), 0.80 \\
(4, 5), 0.67 & (5, 6), 0.92 & (5, 6), 0.92 & (3, 6), 0.92 &  & (3, 9), 0.80& (3, 9), 0.80 \\
(5, 6), 0.67 & (1, 2), 0.67 & (1, 2), 0.87 & (4, 5), 0.92 &  &  & (1, 9), 0.67 \\
(7, 8), 0.67 & (1, 8), 0.67 & (1, 8), 0.87 & (5, 6), 0.92 &  &  &  \\
(7, 9), 0.67 & (7, 9), 0.67 & (7, 9), 0.87 & (1, 2), 0.87 &  &  &  \\
(8, 9), 0.67 & (8, 9), 0.67 & (8, 9), 0.87 & (7, 9), 0.87 &  &  &  \\
(2, 3), 0.00 & (2, 3), 0.50 & (2, 3), 0.70 & (2, 3), 0.70 &  &  &  \\
(3, 7), 0.00 & (3, 7), 0.50 & (3, 7), 0.70 & (3, 7), 0.70 &  &  &  \\
\bottomrule
\end{tabular}
\end{table}

To evaluate the effectiveness and applicability of the proposed cycle similarity metric, we validate it on two classic tasks: link prediction and community detection.

\subsection{Cycle similarity-based link prediction}

Link prediction aims to infer missing or future connections based on the observed structure of a network, with broad applications in social network analysis, recommendation systems, and biological network inference. To assess the performance of link prediction algorithms, we systematically divide the set of existing edges \( E \) into two subsets: a training set \( E_{train} \) and a test set \( E_{test} \). Additionally, we explicitly consider the set of non-existent edges \( E_{non} \). Typically, \( E_{train} \) comprises 90\% of all existing edges, serving as the basis for learning structural patterns and computing node similarity scores. The remaining 10\% of existing edges form \( E_{test} \), which acts as ground truth to evaluate the algorithm's ability to recover missing or potential links. The set \( E_{non} \) includes all possible node pairs that do not exist as edges in the original network. 
Three standard metrics are employed to quantify performance:

\subsubsection{Evaluation metrics}
Area under the receiver operating characteristic curve (AUC) measures the probability that a randomly chosen true edge (from \( E_{test} \)) receives a higher similarity score than a randomly selected non-existent edge (from \( E_{non} \)). It is calculated as: 

\begin{eqnarray}
\text{AUC} = \frac{n' + 0.5n''}{n} 
\end{eqnarray}

where \( n \) is the total number of independent comparisons between a test edge and a non-existent edge,
\( n' \) is the count of comparisons where the test edge has a higher score than the non-existent edge, 
\( n'' \) is the count of comparisons where the test edge and the non-existent edge have equal scores (ties).
An AUC value approaching 1 indicates excellent discriminative power, significantly outperforming random guessing.

Precision evaluates the proportion of correctly predicted edges among the top-L ranked predictions. Here, L is typically set to the size of the test set, i.e., \( L = |E_{test}| \). Formally: 

\begin{eqnarray}
\text{Precision} = \frac{m}{L} 
\end{eqnarray}

where \( m \) is the number of edges in the top-L ranked list that actually belong to \( E_{test} \). Precision focuses on the accuracy of the highest-ranked predictions, making it highly relevant for applications requiring high-confidence recommendations.

Ranking Score assesses the average relative position of test edges within the final ranking of all candidate unseen edges. It is defined as: 

\begin{eqnarray}
\text{Ranking Score} = \frac{1}{|E_{test}|}\sum_{i \in E_{test}} \frac{r_i}{|E_{test}\cup E_{non}|} 
\end{eqnarray}

where \( r_i \) is the rank of the i-th test edge in the ordered list of candidate edges, $|E_{test}|$ is the number of edges in test set.
\( |E_{test}\cup E_{non}| \) is the total number of candidate edges. A lower Ranking Score indicates that true edges (from \( E_{test} \)) are, on average, ranked higher among all candidates, thereby reflecting superior predictive performance.

\subsubsection{Edge-addition correction} 
In many real-world networks, such as tree-like structures or networks with a large number of degree-1 pendant nodes, the number of cycles is often scarce. This cycle sparsity leads to the failure of traditional cycle-based similarity measures (e.g., cycle similarity), as insufficient cycle information hinders the effective capture of deep node associations. To address this issue, we propose an edge-addition correction for cycle similarity. The core idea of this method is to pre-add a virtual edge between two target nodes \( x \) and \( y \) before computing their similarity.
 
Specifically, by introducing a temporary edge between nodes \( x \) and \( y \), we construct new cycle structures that incorporate this edge, thereby enriching local cycle features. Notably, when we set the corrected cycle length threshold to \( L^c_{x,i,y} = 3 \), the modified cycle similarity measure reduces to the classic Common Neighbors (CN) similarity index. This is because, after adding the edge \( (x, y) \), cycles of length 3 essentially correspond to common neighboring nodes \( i \) of \( x \) and \( y \), forming path structures like \( x \rightarrow i \rightarrow y \). Combined with the added edge \( x \rightarrow y \), this forms a triangle. Thus, the process of counting such cycles is equivalent to tallying the number of shared neighbors between nodes \( x \) and \( y \), simplifying the complex cycle structure analysis to intuitive common neighbor counting.
 
This edge-addition correction strategy not only overcomes the failure of traditional cycle similarity in sparse-cycle networks but also enables a smooth transition from local neighborhood structures to global cycle structures by flexibly adjusting the cycle length threshold \( L^c_{x,i,y} \). This enhances the applicability and robustness of the similarity measure across different network types.

\subsection{Cycle similarity-based community detection}
\subsubsection{Cycle-based Hierarchical Clustering (CHC) algorithm}
This algorithm leverages local cyclic patterns to quantify node similarity, capturing higher-order topological features beyond simple pairwise connections, thereby enhancing the detection of communities in networks with rich cyclic structures. 
 
The CHC algorithm operates by first defining node similarity based on shared cyclic structures, then determining inter-community similarity using linkage criteria, and finally performing agglomerative hierarchical clustering while optimizing modularity to identify the optimal community partition. The proposed hierarchical clustering algorithm proceeds as follows:

(1) Initialization: Treat each node as an individual community. Compute the pairwise node similarity matrix \( S \) based on cyclic patterns.

(2) Node Similarity Calculation: For any node pair \( (x, y) \), their similarity \( S(x, y) \) is defined as the sum of the reciprocals of the lengths of the shortest cycles connecting them via shared neighbors, considering only cycles up to a predefined maximum length \( L_{\max}^c \).

(3) Inter-Community Similarity: The similarity between two communities is computed using either the Average (UPGMA) or the Centroid linkage criterion. These rules aggregate the pairwise node similarities to derive a single similarity value for each pair of communities.

(4) Agglomerative Clustering: Iteratively identify and merge the two most similar communities. After each merge, update the inter-community similarity matrix according to the chosen linkage criterion.

(5) Modularity Tracking and Partition Selection: After every merging step, the network modularity \( Q \) is calculated. The partition corresponding to the highest observed modularity during the entire process is retained as the final community structure.

(6) Termination: The procedure continues until all nodes are merged into a single community. The output is the optimal partition identified in step (5).

\subsubsection{Evaluation metrics}
The Fraction of Vertices Identified Correctly (FVIC) measures the proportion of nodes correctly assigned to their ground-truth communities after resolving arbitrary label differences via optimal matching. It is computed as the ratio of correctly matched nodes to the total number of nodes following an optimal one-to-one mapping between predicted and true community labels. In essence, FVIC represents the most intuitive form of accuracy for community assignments. 

The Normalized Mutual Information (NMI) quantifies the agreement or shared information between the detected partition and the ground truth, normalized to a range between 0 (no agreement) and 1 (perfect match). It reflects how well the community structure is recovered, independently of label numbering and robust to differences in the number of communities.

\section{Results and discussions}\label{S3}

\subsection{Link prediction}\label{S3.1}
\subsubsection{Data}
We evaluate link prediction performance on eight real-world networks spanning diverse domains—social, co-occurrence, transportation, collaboration, infrastructure, technological, biological, and political—ensuring a broad assessment of generalizability. Descriptions and key references are summarized below.

The Karate network captures friendship ties among members of a university karate club\cite{Zachary1977}. Miserables is a character co-occurrence network from Victor Hugo's Les Misérables\cite{Miserables1993}. USAir represents the U.S. domestic air transportation network\cite{USAir2006}. NS (Netscience) is a co-authorship network among scientists\cite{NS2006}. Grid models the Western States power grid of the United States\cite{Grid1998}. INT is an autonomous system (AS) level Internet topology\cite{INT2005}. PPI is a protein–protein interaction network\cite{PPI2002}. PB (Political Blogs) records hyperlinks between political blogs during the 2004 U.S. election\cite{PB2005}.

When computing the AUC, it is necessary to sample both the test edge set and the set of non-existent edges. To ensure the statistical reliability of the results, we performed 5,000 independent sampling comparisons. Furthermore, considering the randomness inherent in splitting the data into training and test sets, the reported values of AUC, Precision, and Rank Score are averaged over 100 distinct random splits.

\begin{table}[htbp]
    \centering
    \caption{Basic statistics, motif Z-scores, and common-neighbor link prediction performance of the networks.}
    \label{tab3}
    \begin{tabular}{lccccccccc}
        \toprule
        Network & \textbf{$N$} & \textbf{$M$} & \textbf{$Z_{cycle}^3$} & 
        \textbf{$Z_{cycle}^4$} & \textbf{$Z_{cycle}^5$} & \textbf{CN-AUC} & 
        \textbf{CN-Prec} & \textbf{CN-RS} & \textbf{$N_{k=1}$} \\ 
        \midrule
        karate & 34 & 78 & 6.43 & 1.47 & -0.97 & 0.6746 & 0.1338 & 0.03646 & 1 \\
        Miserables & 77 & 254 & 23.34 & -6.72 & -7.31 & 0.9067 & 0.4832 & 0.00382 & 17 \\
        USAir & 332 & 2126 & 82.62 & -21.55 & -18.67 & 0.9356 & 0.3724 & 0.00041 & 55 \\
        NS & 379 & 914 & 147.95 & -8.29 & -10.95 & 0.9502 & 0.3691 & 0.00034 & 27 \\
        Grid & 4941 & 6594 & 371.86 & 138.39 & 56.31 & 0.5881 & 0.0430 & 0.00066 & 1226 \\
        INT & 5022 & 6258 & 21.41 & 81.76 & -5.01 & 0.5590 & 0.0172 & 0.00074 & 3259\\
        PPI & 2375 & 11693 & 410.17 & 23.01 & -59.09 & 0.8932 & 0.1439 & 0.00008 & 505 \\
        PB & 1222 & 19021 & 41.85 & -25.31 & -51.97 & 0.9185 & 0.1718 & 0.00004 & 127 \\
        \bottomrule
    \end{tabular}
\end{table}

Table \ref{tab3} presents partial statistical properties of these networks, where $N$ denotes the number of nodes, $M$ represents the number of edges, $Z_{\text{cycle}}^3$, $Z_{\text{cycle}}^4$, and $Z_{\text{cycle}}^5$ indicate the Z-scores for three-node, four-node, and five-node cycles, respectively; CN-AUC, CN-Prec, and CN-RS correspond to the AUC value, precision value, and rank score measured via the common neighbor similarity index. It should be noted that the Z-score originates from motif analysis and is defined as the difference between the occurrence count of a motif in the real network and its mean occurrence in randomized networks, divided by the standard deviation of occurrences in the randomized ensemble \cite{Science2002,NRG2007,Science2016}. It quantifies the statistical significance of the motif relative to random networks. A larger absolute Z-score indicates higher significance of the motif in the real network; a positive value suggests that the motif appears more frequently than in random networks, whereas a negative value implies it appears less frequently.

\subsubsection{Basic cycle similarity}
Table \ref{tab4} describes the link prediction metrics measured under the constraints of maximum cycle sizes \(L^c_{max}=4\) and \(L^c_{max}=5\). The results show that the AUC values obtained across all datasets are slightly lower than those derived from the common neighbor similarity index. Except for the USAir and Miserables datasets, the AUC values for \(L^c_{max}=5\) are marginally higher than those for \(L^c_{max}=4\). Regarding Precision, except for the INT and PPI datasets, the values for \(L^c_{max}=5\) are lower than those for \(L^c_{max}=4\). Changes in Rank score values are relatively subtle, making it difficult to discern clear distinctions.

\begin{table}
	\centering 
	\caption{Comparative performance of link prediction metrics with varying cycle size limits.}
	\label{tab4}
	\begin{tabular}{lccc|ccc}
		\toprule 
		\multirow{2}{*}{Network} & \multicolumn{3}{c|}{$L^c_{max}=4$} & \multicolumn{3}{c}{$L^c_{max}=5$} \\
		\cmidrule(lr){2-4} \cmidrule(lr){5-7}
		& \textbf{AUC} & \textbf{Prec.} & \textbf{RS} & \textbf{AUC} & \textbf{Prec.} & \textbf{RS} \\ 
		\midrule 
		karate & 0.6089 & 0.1513 & 0.05678 & 0.6545 & 0.1338 & 0.04666 \\
        Miserables & 0.8897 & 0.4556 & 0.00630 & 0.8883 & 0.2344 & 0.00475 \\
        USAir & 0.9202 & 0.3720 & 0.00041 & 0.9133 & 0.2705 & 0.00051 \\
        NS & 0.8313 & 0.3615 & 0.00151 & 0.8507 & 0.2323 & 0.00187 \\
        Grid & 0.5212 & 0.0428 & 0.00070 & 0.5290 & 0.0356 & 0.00067 \\
        INT & 0.5311 & 0.0160 & 0.00076 & 0.5544 & 0.0844 & 0.00065 \\
        PPI & 0.8447 & 0.1448 & 0.00018 & 0.8755 & 0.2302 & 0.00016 \\
        PB & 0.9036 & 0.1727 & 0.00005 & 0.9159 & 0.1046 & 0.00007 \\
		\bottomrule 
	\end{tabular}
\end{table}

\subsubsection{Edge-added corrected cycle similarity}
Building upon the discussion of cycle similarity, its limitations—such as computational complexity and suboptimal AUC results, particularly in sparse networks where cycles are scarce—motivate a refined approach. To enhance its effectiveness, we introduce an edge-addition correction. This method enriches local cycle information by temporarily adding a virtual edge between two target nodes before similarity computation. Notably, setting the corrected cycle length threshold to three reduces this modified measure exactly to the common neighbors index, thereby bridging a well-established baseline with the cycle-based framework.

\begin{table}
	\centering 
	\caption{Performance comparison of the edge-corrected cycle similarity under varying $L^c_{max}$ settings against classical similarity indices.}
	\label{tab5}
	\begin{tabular}{lccc|ccc|ccc}
		\toprule 
		\multirow{2}{*}{Network} & \multicolumn{3}{c|}{$L^c_{max}=3$} & \multicolumn{3}{c|}{$L^c_{max}=4$} & \multicolumn{3}{c}{$L^c_{max}=5$} \\
		\cmidrule(lr){2-4} \cmidrule(lr){5-7} \cmidrule(lr){8-10}
		& \textbf{AUC} & \textbf{Prec.} & \textbf{RS} & \textbf{AUC} & \textbf{Prec.} & \textbf{RS} & \textbf{AUC} & \textbf{Prec.} & \textbf{RS} \\ 
		\midrule 
		karate & 0.6702 & 0.1475 & 0.04250 & 0.7779 & 0.1500 & 0.02446 & 0.7657 & 0.0900 & 0.03201 \\
        Miserables & 0.9023 & 0.4752 & 0.00384 & 0.8505 & 0.2328 & 0.00571 & 0.8163 & 0.1684 & 0.00755 \\
        USAir & 0.9356 & 0.3733 & 0.00023 & 0.9021 & 0.3596 & 0.00034 & 0.8847 & 0.2904 & 0.00038 \\
        NS & 0.9478 & 0.3319 & 0.00065 & 0.9466 & 0.1901 & 0.00078 & 0.9307 & 0.1196 & 0.00070 \\
        Grid & 0.5883 & 0.0427 & 0.00071 & 0.6388 & 0.0355 & 0.00053 & 0.6801 & 0.0293 & 0.00053 \\
        INT & 0.5598 & 0.0156 & 0.00070 & 0.6343 & 0.0874 & 0.00065 & 0.6170 & 0.0565 & 0.00061 \\
        PPI & 0.8921 & 0.1435 & 0.00008 & 0.9393 & 0.2538 & 0.00006 & 0.9208 & 0.2331 & 0.00008 \\
        PB & 0.9185 & 0.1727 & 0.00004 & 0.9269 & 0.1332 & 0.00003 & 0.9004 & 0.0883 & 0.00008\\
		\bottomrule 
	\end{tabular}
\end{table}

Table \ref{tab5} presents the link prediction results after applying the edge-addition correction. It can be observed that with \(L^c_{max}=3\), the results are essentially consistent with those obtained from the common neighbor (CN) similarity index. This equivalence stems from the fact that adding an auxiliary edge between nodes \(x\) and \(y\) allows them to form triangles with each common neighbor, making the count of three-node cycles identical to the count of common neighbors. Interestingly, after introducing the auxiliary edge, the AUC values improve for most datasets. In particular, for the INT, PPI, Grid, and Karate networks, the AUC increases further as \(L^c_{max}\) grows, eventually outperforming classical node similarity indices such as Resource Allocation (RA) and Adamic-Adar (AA). This enhancement can likely be attributed to the fact that adding the auxiliary edge facilitates the formation of cycles in the network. In this setting, searching for the shortest cycle containing the added edge is essentially equivalent to finding the shortest path between the neighbors of the source node and the other target node after removing that source node, thereby capturing deeper node associations more effectively.

\subsection{Community detection}\label{S3.2}

\subsubsection{Data}
To assess community detection performance, we employ eight real-world networks commonly used as benchmarks, covering social, organizational, informational, and communication systems. Brief descriptions are as follows.

The Karate network records friendships within a university club, famously split into two factions\cite{Zachary1977}. The Dolphins network maps frequent associations within a dolphin community\cite{Dolphins2003}. The Football network represents games between U.S. college teams, organized into conferences\cite{Football2002}. The Polbooks network links politically-themed books based on co-purchasing patterns\cite{polbooks2004}. The Prischool1 and Prischool2 networks capture weighted, face-to-face contact patterns in a primary school, where communities correspond to classrooms\cite{Prischool2011}. The Polblogs network is a directed hyperlink graph of political blogs, naturally divided by ideology\cite{PB2005}. Lastly, the Email network models internal email exchanges within a research institution, revealing its functional organization\cite{Email2009}.

\begin{table}[htbp]
    \centering
    \caption{Statistical and cyclic properties of the datasets for community detection.}
    \label{tab6}
    \begin{tabular}{lcccccccc}
        \toprule
        Network & \textbf{$N$} & \textbf{$M$} & \textbf{$C_n$} & 
        \textbf{$Q$} & \textbf{$Z_{cycle}^3$} & \textbf{$Z_{cycle}^4$} & 
        \textbf{$Z_{cycle}^5$} & \textbf{$N_{k=1}$} \\ 
        \midrule
        karate & 34 & 78 & 2 & 0.3582 & 5.23 & 1.96 & -0.65 & 1 \\
        dolphins & 62 & 159 & 2 & 0.3735 & 13.93 & -1.4 & -1.94 & 9 \\
        football & 115 & 613 & 12 & 0.5540 & 60.60 & -6.25 & -13.06 & 0 \\
        polbooks & 105 & 441 & 2 & 0.3951 & 31.64 & -2.08 & -10.13 & 0 \\
        Prischool1 & 236 & 5899 & 11 & 0.2930 & 92.41 & -13.03 & -41.28 & 0 \\
        Prischool2 & 238 & 5539 & 11 & 0.3237 & 112.42 & -19.04 & -36.86 & 0 \\
        polblogs & 1222 & 16714 & 2 & 0.4052 & 100.74 & 8.58 & -35.29 & 135 \\
        Email & 986 & 16687 & 42 & 0.3130 & 129.72 & -11.35 & -38.89 & 78 \\
        \bottomrule
    \end{tabular}
\end{table}

Table \ref{tab6} presents partial statistical and cyclic properties of the datasets used for community detection tasks, including the number of nodes \(N\), the number of edges \(M\), the number of ground-truth communities \(C_n\) and the corresponding modularity \(Q\), the Z-scores for three-, four-, and five-node cycles (denoted as \(Z_{\text{cycle}}^3\), \(Z_{\text{cycle}}^4\), \(Z_{\text{cycle}}^5\)), and the number of pendant nodes with degree one, \(N_{k=1}\).

\subsubsection{Hierarchical community detection process}

\begin{figure}
	\centering
	\includegraphics[width=14cm]{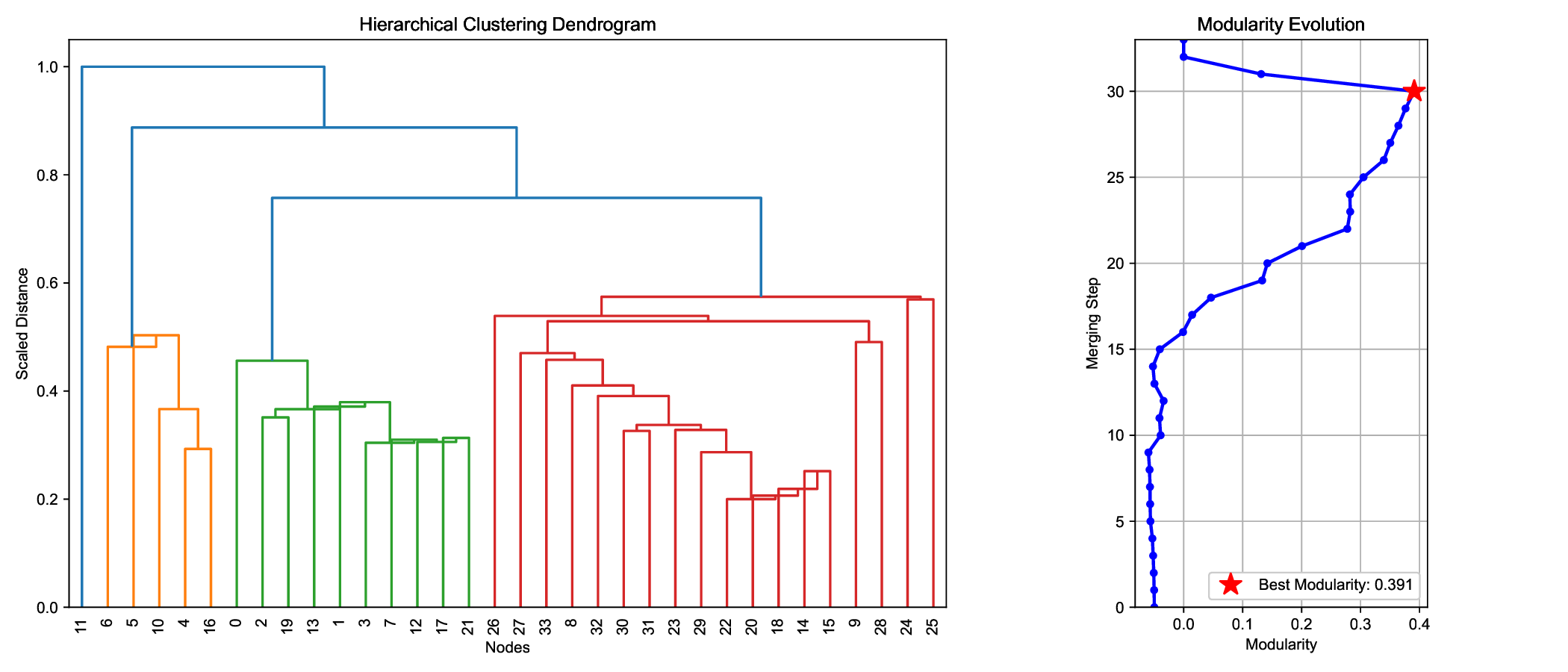}
	\caption{Hierarchical clustering dendrogram and modularity progression for the Karate Club network.}\label{fig2}
\end{figure}

\begin{figure}
	\centering
	\includegraphics[width=12cm]{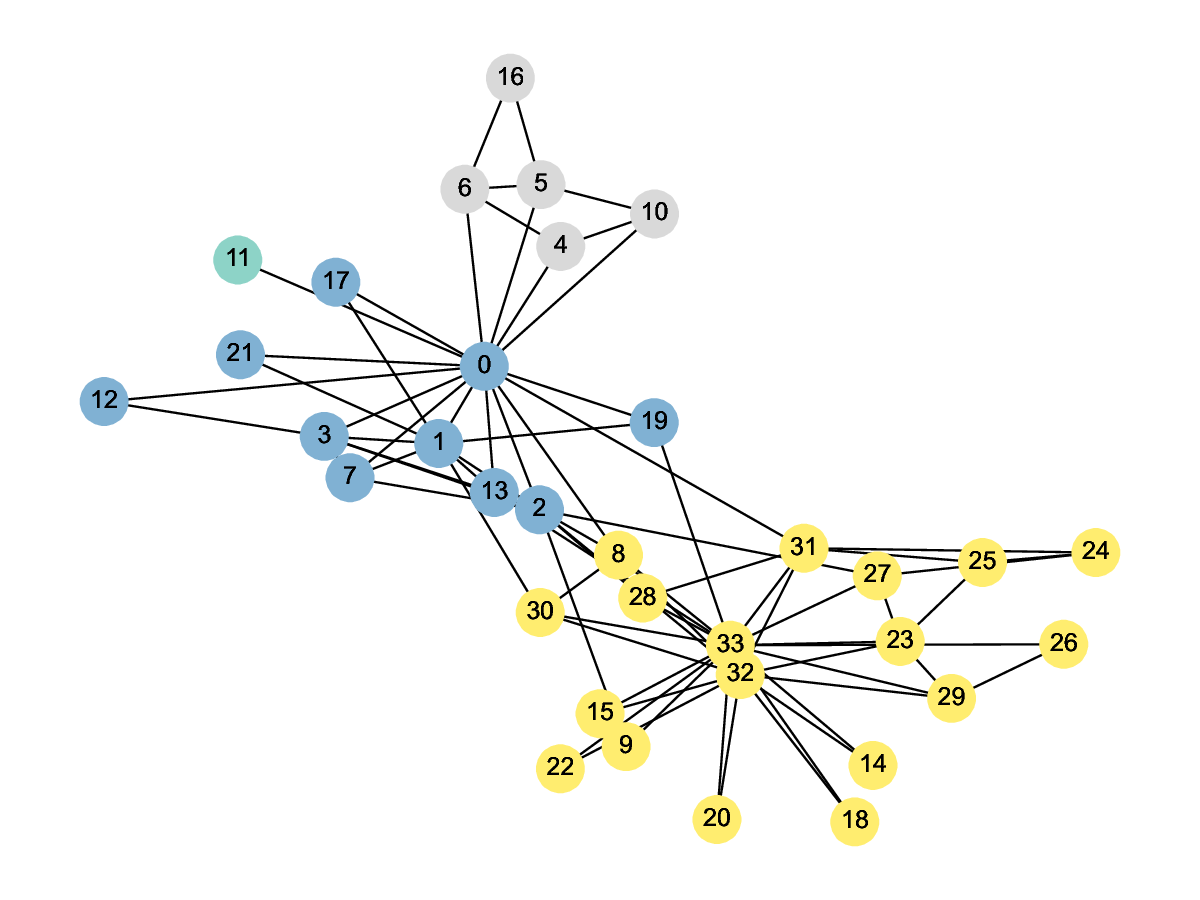}
	\caption{Visualization of the detected community structure in the Karate Club network.}\label{fig3}
\end{figure}

To clearly demonstrate the proposed cycle-similarity-based hierarchical community detection algorithm, this section illustrates its application using the widely studied Karate Club network. Figure \ref{fig2} displays the dendrogram generated during the hierarchical clustering process and the corresponding changes in network modularity at each merging step. It should be noted that the centroid method was used to measure similarity between communities, which explains the occasional decreases observed in the merging trajectory. The dendrogram suggests that the network can be broadly divided into four communities. Specifically, node 11 forms a singleton community and remains isolated until the final merging step. The maximum modularity is achieved at the fourth-to-last step; therefore, following the principle of modularity maximization, the network should be partitioned into four communities. Figure \ref{fig3} visualizes these four communities, each represented by a distinct color. It can be clearly observed that node 11, with a degree of one, is a pendant node that does not participate in any cycle, making it difficult to assign to other communities. Another small community consists of nodes 4, 5, 6, 10, and 16, which connect to the main network through node 0. In network terms, node 0 acts as a cut vertex; its removal would disconnect this small group from the main structure. The two larger communities are respectively centered around node 0 and node 33, who represent the instructor and the club administrator in the original social context.

\subsubsection{Evaluation of community detection results}
Tables \ref{tab7}, \ref{tab8}, and \ref{tab9} present the results of hierarchical community detection based on cycle similarity under three different maximum cycle size settings (\(L^c_{max}=3, 4, 5\)), listing the corresponding Normalized Mutual Information (NMI), Fraction of Vertices Identified Correctly (FVIC), optimal modularity \(Q\) obtained during the merging process, and the associated number of communities \(C_n\) for both the centroid and average methods used to measure inter-community similarity. The results indicate that the best merging performance is achieved with \(L^c_{max}=3\) for the majority of networks, which aligns with the cycle Z-score patterns shown in Table 6—most networks exhibit higher Z-scores for three-node cycles, while only the karate and polblogs datasets show positive Z-scores for four-node cycles. Furthermore, it can be observed from the tables that networks with better hierarchical community detection performance generally possess larger cycle Z-scores and do not contain an excessive number of pendant nodes with degree one.

\begin{table}[htbp]
    \centering
    \caption{Hierarchical community detection performance using cycle similarity with $L^c_{max}=3$}
    \label{tab7}
    \begin{tabular}{lcccccccc}
        \toprule
        Network & \textbf{Ctr\_NMI} & \textbf{Ctr\_FVIC} & \textbf{Ctr\_Cn} & 
        \textbf{Ctr\_Q} & \textbf{Avg\_NMI} & \textbf{Avg\_FVIC} & 
        \textbf{Avg\_Cn} & \textbf{Avg\_Q} \\ 
        \midrule
        karate & 0.7229 & 0.9118 & 4 & 0.3648 & 0.5453 & 0.6471 & 6 & 0.4047 \\
        dolphins & 0.4298 & 0.4355 & 20 & 0.4364 & 0.3911 & 0.3548 & 23 & 0.4032 \\
        football & 0.8561 & 0.8000 & 9 & 0.6044 & 0.8212 & 0.7304 & 8 & 0.6018 \\
        polbooks & 0.4661 & 0.7810 & 6 & 0.5116 & 0.4557 & 0.7333 & 8 & 0.5180 \\
        Prischool1 & 0.8079 & 0.6398 & 7 & 0.3669 & 0.7394 & 0.4873 & 5 & 0.3703 \\
        Prischool2 & 0.5654 & 0.3992 & 5 & 0.3552 & 0.7273 & 0.5000 & 7 & 0.3935 \\
        polblogs & 0.4070 & 0.7750 & 238 & 0.4141 & 0.2897 & 0.5679 & 473 & 0.3844 \\
        Email & 0.6378 & 0.4209 & 234 & 0.3376 & 0.6681 & 0.4057 & 383 & 0.2903 \\
        \bottomrule
    \end{tabular}
\end{table}

\begin{table}[htbp]
    \centering
    \caption{Hierarchical community detection performance using cycle similarity with $L^c_{max}=4$}
    \label{tab8}
    \begin{tabular}{lcccccccc}
        \toprule
        Network & \textbf{Ctr\_NMI} & \textbf{Ctr\_FVIC} & \textbf{Ctr\_Cn} & 
        \textbf{Ctr\_Q} & \textbf{Avg\_NMI} & \textbf{Avg\_FVIC} & 
        \textbf{Avg\_Cn} & \textbf{Avg\_Q} \\ 
        \midrule
        karate & 0.6529 & 0.7941 & 4 & 0.3911 & 0.5653 & 0.7059 & 6 & 0.3625 \\
        dolphins & 0.4136 & 0.3871 & 20 & 0.4288 & 0.5030 & 0.6613 & 16 & 0.4185 \\
        football & 0.8627 & 0.8087 & 9 & 0.6023 & 0.8506 & 0.8000 & 9 & 0.6042 \\
        polbooks & 0.4215 & 0.7333 & 6 & 0.5131 & 0.4600 & 0.7429 & 6 & 0.5148 \\
        Prischool1 & 0.6598 & 0.5424 & 17 & 0.2930 & 0.6318 & 0.4915 & 9 & 0.3187 \\
        Prischool2 & 0.6025 & 0.5126 & 8 & 0.3283 & 0.6059 & 0.5084 & 12 & 0.3237 \\
        polblogs & 0.4757 & 0.8363 & 160 & 0.4188 & 0.3098 & 0.6244 & 343 & 0.3924 \\
        Email & 0.6280 & 0.2961 & 471 & 0.2213 & 0.6203 & 0.3316 & 387 & 0.2777 \\
        \bottomrule
    \end{tabular}
\end{table}

\begin{table}[htbp]
    \centering
    \caption{Hierarchical community detection performance using cycle similarity with $L^c_{max}=5$}
    \label{tab9}
    \begin{tabular}{lcccccccc}
        \toprule
        Network & \textbf{Ctr\_NMI} & \textbf{Ctr\_FVIC} & \textbf{Ctr\_Cn} & 
        \textbf{Ctr\_Q} & \textbf{Avg\_NMI} & \textbf{Avg\_FVIC} & 
        \textbf{Avg\_Cn} & \textbf{Avg\_Q} \\ 
        \midrule
        karate & 0.5135 & 0.7353 & 4 & 0.3339 & 0.2374 & 0.5882 & 4 & 0.1457 \\
        dolphins & 0.4651 & 0.6452 & 14 & 0.3951 & 0.4855 & 0.6452 & 18 & 0.4065 \\
        football & 0.8347 & 0.7478 & 12 & 0.5779 & 0.9028 & 0.8348 & 11 & 0.6012 \\
        polbooks & 0.4760 & 0.8190 & 4 & 0.4864 & 0.4498 & 0.7619 & 6 & 0.5182 \\
        Prischool1 & 0.5697 & 0.2034 & 134 & 0.1112 & 0.5680 & 0.1822 & 139 & 0.0917 \\
        Prischool2 & 0.4856 & 0.3025 & 49 & 0.2409 & 0.5603 & 0.1807 & 133 & 0.1452 \\
        \bottomrule
    \end{tabular}
\end{table}

\section{Conclusion}\label{S4}
Given the prevalence of cyclic structures in real-world networks, this paper proposes a node similarity metric based on minimal cycles, which quantifies the similarity between two nodes by enumerating the minimal cycles connecting them via their neighboring nodes, with an imposed upper bound on cycle size to ensure computational tractability. We then investigate the application of this similarity measure in two fundamental tasks: link prediction and community detection. To address the issue of cycle sparsity in link prediction, an edge-addition correction strategy is introduced. Experimental results demonstrate that this correction leads to improved performance on datasets such as karate, INT, PPI, and Grid. For community detection, we propose a hierarchical clustering algorithm based on cycle similarity. Our findings indicate that the presence of pendant nodes (degree = 1) and cut vertices are the two primary factors limiting the algorithm's performance. Better community detection results are generally achieved when the network exhibits higher cycle Z-scores and contains fewer pendant nodes.

\section*{Data Availability Statement}
The data used in the creation of this manuscript will be made available upon request.

\section*{Conflict of interest}
The authors declare that there is no conflict of interests between them.

\end{document}